\documentclass[aps,prd,showpacs,amsmath,nofootinbib,
superscriptaddress,twocolumn,preprintnumbers]{revtex4}
\usepackage{natbib}
\usepackage{amsmath}
\usepackage{amssymb}
\usepackage{graphicx}
\usepackage{color}
\usepackage{amsmath}
\usepackage{amsthm}
\usepackage{slashed}
\usepackage{soul}
\begin{document}

\title{Homogeneous and Isotropic Spacetime in Conformal Scalar-Tensor Gravity}

\author{Meir Shimon}
\affiliation{School of Physics and Astronomy, 
Tel Aviv University, Tel Aviv 69978, Israel}
\email{meirs@wise.tau.ac.il}

\begin{abstract}
The background field equations for homogeneous and isotropic spacetime are derived in 
conformal scalar-tensor gravity. The background temporal evolution is entirely driven 
by the dynamical evolution of the scalar field, i.e. particle masses, and satisfies an 
equation which is identical in form to the Friedmann equation of the standard cosmological 
model in general relativity. In a static background spacetime the scalar field (logarithmic) 
time-derivative replaces the `Hubble function'. It is also shown that linear 
perturbations are governed by equations which are identical to those 
obtained in general relativity, but with 
their evolution stemming from the scalar field dynamics.
\end{abstract}


\maketitle

\section{Introduction}

A non-vacuum homogeneous and isotropic spacetime is described in 
general relativity (GR) by the dynamical Friedmann-Robertson-Walker 
(FRW), which can be either expanding or contracting. In this theoretical framework 
redshift of light from distant objects is commonly interpreted as evidence for space 
expansion.

The main objective of the present work is to demonstrate that 
in the framework of a conformal scalar-tensor gravity [1] temporal evolution 
of the scalar field in homogeneous and isotropic {\it static} space can provide 
a viable alternative to space expansion: 
Cosmological redshift in particular, and cosmic evolution in general, 
are driven 
by the time evolution of the monotonically increasing masses of fundamental particles.

We work in a mostly-positive metric signature convention, and adopt units 
such that $\hbar=1=c$. Our theoretical approach is outlined 
in section II, and the model describing a 
maximally-symmetric static spacetime is presented in section 
III, and summarized in section IV.

\section{Theoretical Framework}

The theoretical framework adopted here is formally in the class 
of generalized Brans-Dicke (BD) theories 
where the matter lagrangian explicitly depends on the scalar field, 
(as in Bergmann-Wagoner gravity [2, 3]), i.e. both the {\it effective} 
gravitational `constant' $G$ and particle masses are 
spacetime-dependent. Specified to inertial/gravitational 
phenomena, the action described in [1] reduces to 
\begin{eqnarray}
\mathcal{I}_{CSM}&=&\int\left[\frac{1}{6}|\phi|^{2}R
+\phi_{\mu}^{*}\phi^{\mu}-\lambda|\phi|^{4}\right.\nonumber\\
&+&\left.\mathcal{L}_{M}(\phi,\phi^{*},g_{\mu\nu})\right]\sqrt{-g}d^{4}x. 
\end{eqnarray} 
Here, $\phi$ and $g_{\mu\nu}$ are the scalar and metric fields 
respectively, $\lambda$ is a dimensionless self-coupling parameter, 
the ordinary derivative of $\phi$ with respect to $x^{\mu}$ is 
denoted $\phi_{\mu}\equiv \phi_{,\mu}$. 
The lagrangian density of the non-pure-scalar field sector, $\mathcal{L}_{M}$,  
is assumed to depend on $\phi$ via Yukawa-like couplings of 
the form $\lambda_{y,i}\bar{\psi}\phi\psi+h.c.$ where $\lambda_{y,i}$ is the 
dimensionless Yukawa-like coupling of the i'th particle species, i.e. 
particle masses are proportional to $|\phi|$, in analogy with the standard 
model of particle physics. This lagrangian 
describes gravitational and inertial phenomena [1], 
indeed the main interest of the current work. 
Eq. (1) is conformally-invariant under {\it local} re-scaling, 
$\phi\rightarrow\phi\Omega^{-1}$
and $g_{\mu\nu}\rightarrow g_{\mu\nu}\Omega^{2}$, 
where $\Omega(x)$ is an arbitrary function of spacetime. 
The action for a pointlike massive particle 
$\mathcal{L}_{p}\propto|\phi|\sqrt{\int g_{\mu\nu}\frac{dx^{\mu}}{d\xi}
\frac{dx^{\nu}}{d\xi}d\xi}$ is similarly invariant under this transformation, 
where $\xi$ is the affine parameter.

Variation of Eq. (1) with respect to $g_{\mu\nu}$ and $\phi^{*}$ results in 
the generalized Einstein equations, and scalar field equation, respectively, 
\begin{eqnarray}
\frac{|\phi|^{2}G_{\mu}^{\nu}}{3}&=&T_{M,\mu}^{\nu}
-\Theta_{\mu}^{\nu}-\lambda\delta_{\mu}^{\nu}|\phi|^{4}\\
\frac{\phi R}{6}-\Box\phi&-&2\lambda|\phi|^{2}\phi
+\frac{\partial\mathcal{L}_{M}}{\partial\phi^{*}}=0,
\end{eqnarray}
and the generalized energy momentum (non-) conservation 
then follows, e.g. [4, 5]
\begin{eqnarray}
T_{M,\mu;\nu}^{\nu}&=&\mathcal{L}_{M,\phi}\phi_{\mu}+\mathcal{L}_{M,\phi^{*}}\phi^{*}_{\mu}.
\end{eqnarray}
The effective energy-momentum tensor $\Theta_{\mu}^{\nu}$ 
associated with the scalar field is
\begin{eqnarray}
\Theta_{\mu}^{\nu}&=&\frac{1}{3}\delta_{\mu}^{\nu}(\phi^{*}\Box\phi+\phi\Box\phi^{*}
-\phi^{*}_{\rho}\phi^{\rho})\nonumber\\
&+&\frac{1}{3}(2\phi^{*}_{\mu}\phi^{\nu}+2\phi_{\mu}\phi^{*\nu}
-\phi^{*}\phi_{\mu}^{\nu}-\phi\phi_{\mu}^{*\nu}).
\end{eqnarray}
Here and throughout, $f_{\mu}^{\nu}\equiv(f_{,\mu})^{;\nu}$, with $f_{;\mu}$ 
denoting covariant derivatives of $f$, the covariant Laplacian is $\Box f$, 
and $(T_{M})_{\mu\nu}\equiv -\frac{2}{\sqrt{-g}}\frac{\delta(\sqrt{-g}\mathcal{L}_{M})}
{\delta g^{\mu\nu}}$ is the energy-momentum tensor.
Eq. (5), which is not independent of (2) \& (3), 
implies that energy-momentum (of matter alone) is generally not 
conserved, which is expected in the case that $\Lambda$ or particle masses are 
spacetime-dependent. Finally, combining Eq. (3) with the trace of Eq. (2), 
we obtain the following consistency relation
\begin{eqnarray}
\phi^{*}\frac{\partial\mathcal{L}_{M}}{\partial\phi^{*}}
+\phi\frac{\partial\mathcal{L}_{M}}{\partial\phi}=T_{M}.
\end{eqnarray}

\section{Background Equations and Linear Perturbations}

We describe the background evolution and the evolution of linear 
perturbations in a homogeneous and isotropic static background 
spacetime and show equivalence with the standard dynamics of the corresponding 
FRW solution in GR and its linear perturbations.

\subsection{Evolution of the Background Fields}

The background evolution of the scalar field 
(to which particle masses are proportional [1]) 
is described on a {\it static} maximally-symmetric background spacetime. 
The Einstein tensor components $G_{\mu}^{\nu}$, associated with 
the metric $g_{\mu\nu}=diag(-1,\frac{1}{1-Kr^{2}},r^{2},r^{2}\sin^{2}\theta)$,
with conformal rather than cosmic time, are $G_{\eta}^{\eta}=-3K$ and 
$G_{i}^{j}=-K\delta_{i}^{j}$.
Here, `$i,j$' indices stand for the spatial coordinates. 

The energy-momentum tensor of a perfect fluid is 
$(T_{M})_{\mu}^{\nu}=\rho_{M}\cdot diag(-1,w_{M},w_{M},w_{M})$, 
where $w_{M}=P_{M}/\rho_{M}$ is the equation of state (EOS) 
describing a matter with energy density $\rho_{M}$ and pressure $P_{M}$.
In case that $\mathcal{L}_{M}=-\rho_{M}$ 
(i.e., that $\mathcal{L}_{M}$ does not explicitly depend on the scalar field 
derivatives) then it immediately follows from Eq. (6) 
that $\rho_{M}\propto|\phi|^{1-3w_{M}}$ where $w_{M}\neq 0$. 
The case $w_{M}=0$ is treated separately below. 
As expected, $\rho_{M}$ is a quartic potential in the case $w_{M}=-1$, 
and is independent of $|\phi|$, i.e. of masses, in the case $w_{M}=1/3$.
In the special case of nonrelativistic (NR) fermions ($w_{M}=0$) that 
are described by the 
Yukawa-like term implicitly appearing in $\mathcal{L}_{M}$ [1], 
it readily follows from Eq. (6) in flat and 
static background metric, i.e. $\rho'_{M}=\lambda_{y,i}\bar{\psi}\psi\phi'+h.c.
=\frac{1}{2}\rho_{M}\left(\phi'/\phi+h.c.\right)$ 
where $\bar{\psi}\psi=constant$ (since $\bar{\psi}\psi$ 
is proportional to the number density of NR fermionic 
particles in static space), that $\rho_{M}\propto |\phi|$. Together 
with the result $\rho_{M}\propto|\phi|^{1-3w_{M}}$ 
for $w_{M}\neq 0$ we conclude that $\rho_{M}\propto |\phi|^{1-3w_{M}}$ 
for any $w_{M}$. 
Here, $f'\equiv\frac{\partial f}{\partial\eta}$ is the derivative 
of a function $f$ with respect to conformal time $\eta$.

In the following we consider a real scalar field, 
i.e. $\phi=|\phi|e^{i\theta}$, with $\theta=0$, and for notational simplicity 
we denote the scalar field modulus $\phi$ where it is clear that $\phi>0$. 
In the full cosmological model [6] we relax this 
and indeed show that 
the phase $\theta$ plays a crucial role.
Using $\rho_{M}\propto\phi^{1-3w_{M}}$ Eqs. (3) become
\begin{eqnarray}
\mathcal{Q}^{2}+K&=&\frac{\rho_{M}}{\phi^{2}}\\
2\mathcal{Q}'+\mathcal{Q}^{2}+K&=&-\frac{3w_{M}\rho_{M}}{\phi^{2}},
\end{eqnarray}
where we used $\mathcal{Q}\equiv\phi'/\phi$. 
In comparison, the standard Friedmann equation 
reads $\mathcal{H}^{2}+K=a^{2}(\eta)\rho_{M}$ 
in conformal time coordinates where $a(\eta)$ is the scale factor 
(rescaled by a factor $\sqrt{\frac{8\pi G}{3}}$), $\mathcal{H}\equiv a'/a$, 
and $\rho_{M}\propto a^{-3(1+w_{M})}$, i.e. the right hand side of 
the Friedmann equation is $\propto a^{-1-3w_{M}}$. 

This is analogous to the term $\rho_{M}/\phi^{2}\propto\phi^{-1-3w_{M}}$ 
appearing on the right hand side of Eq. (7) when 
$a(\eta)$ is replaced with $\phi(\eta)$, i.e. $\mathcal{H}$ is replaced 
by $\mathcal{Q}$ throughout. Consequently, Eqs. (7) \& (8) 
are identical in form to the Friedmann equations in conformal time coordinates. 
In other words, the relation between $\eta$ and the energy-density, i.e. 
the distance-redshift relation, is unchanged compared to standard cosmology.
We thus obtained an equivalent description to that of the standard cosmological 
model in conformal scalar-tensor gravity, at least at the background level. 

The energy-momentum tensor of matter itself is known to be generally non-conserved 
in scalar-tensor theories of gravity, as is indeed evident from Eq. (4). 
By virtue of the formal equivalence to the Friedmann equation 
in the standard cosmological model, the conserved 
quantity in the homogeneous and isotropic background is $\rho_{M}/\phi^{4}$. Another way 
to see this is to consider the continuity equation, 
which reads $\rho'_{M}+3(1+w_{M})\mathcal{H}\rho_{M}=0$ in FRW spacetimes. 
The analogous (non-) conservation equation in a static background model 
is Eq. (4), which yields 
$\left(\frac{\rho_{M}}{\phi^{4}}\right)'
+3(1+w_{M})\mathcal{Q}\left(\frac{\rho_{M}}{\phi^{4}}\right)=0$,  
and $\rho_{M}\propto\phi^{1-3w_{M}}$ is recovered. 
This is analogous to the relation obtained from 
the Friedmann equation, $\phi(\eta)\propto\eta^{2/(1+3w_{M})}$. 

In the redshifting universe $\phi$ is a monotonically increasing function of 
conformal time for any $w_{M}>-1/3$. 
Thus, unlike in the standard FRW spacetime, the observed cosmological redshift 
is not due to stretching of photon wavelengths in an expanding background, but rather 
due to evolving particle masses, e.g., 
due to the evolving Rydberg `constant' on a {\it static} background space. 
What is normally considered a {\it static} 
object in standard cosmology, e.g. a galaxy, star, etc., is replaced in the present work 
by a {\it stationary} solution. More specifically, Weyl transformation is applied 
to any static solution $\phi(r)$ \& $g_{\mu\nu}(r)$ 
with $\Omega(\eta)\propto\eta^{-2/(1+3w_{M})}$ so as to guarantee that particle masses 
transform continuously between the background and objects residing in it, e.g. 
galaxies, stars, etc, exactly 
as (fixed) masses (trivially) transform continuously in standard cosmology between 
expanding background space and compact objects; mass is fixed in standard cosmology 
{\it by construction}. 
Thus, the ratio between inertial and metric length scales, 
$\phi^{-1}/\sqrt{g_{\mu\nu}}$, is still static. From this perspective, observers 
in shrinking galaxies and stars residing in a static background would observe 
a redshifting universe.  

\subsection{Linear Perturbation Theory}

As in section III.A, we assume an effective single fluid with 
matter density $\rho_{M}$ and a (generally time-dependent) $w_{M}=w_{M}(\eta)$. 
In the following we show that the peculiar velocity of matter, $v$,
and the effective $\alpha$, $\varphi$, $\delta_{\rho_{M}}$, and $\delta_{P_{M}}$, 
evolve in the same fashion as in standard cosmology. Here, $\varphi$ and $\alpha$ 
are the Newtonian and curvature gravitational potentials, respectively, 
appearing in the perturbed FRW line element
$ds^{2}=\phi^{2}[-(1+2\alpha)d\eta^{2}+(1+2\varphi)\gamma_{ij}dx^{i}dx^{j}]$ 
where $\gamma_{ij}\equiv diag[1/(1-Kr^{2}),r^{2},r^{2}\sin^{2}\theta]$, 
and Latin indices here run over space coordinates.
The energy density and pressure perturbations in energy density units 
are $\delta_{\rho_{M}}\equiv\delta\rho_{M}/\rho_{M}$ and 
$\delta_{P_{M}}\equiv\delta P_{M}/\rho_{M}$, respectively. 
The anisotropic stress is also redefined, $\pi^{(s)}\rightarrow \phi^{4}\pi^{(s)}$.

Next we write down the dynamical equations governing the evolution of 
metric and matter perturbations in the shear-free gauge. 
The linear order scalar perturbation equations are summarized in [4]. 
Defining the `shifted' quantities $\alpha\rightarrow\alpha-\delta_{\phi}$, 
$\varphi\rightarrow\varphi-\delta_{\phi}$, 
$\delta_{\rho_{M}}\rightarrow\delta_{\rho_{M}}+4\delta_{\phi}$ 
and $\delta_{P_{M}}\rightarrow\delta_{P_{M}}+4\delta_{\phi}$, 
where $\delta_{\phi}\equiv\delta\phi/\phi$, 
we obtain for the Arnowitt-Deser-Misner (ADM) 
energy (time-time), momentum (time-space), 
and propagation (space-space) components, 
(Eqs. 38, 40 \& 41 of [4], respectively)
\begin{eqnarray}
&&(k^{2}/3-K)\varphi-\mathcal{Q}^{2}\alpha
+\mathcal{Q}\varphi'=\frac{1}{2}(\mathcal{Q}^{2}+K)\delta_{\rho_{M}}\\
&&-\varphi'+\mathcal{Q}\alpha=\frac{3(1+w_{M})}{2}(\mathcal{Q}^{2}+K)u\\
&&\alpha+\varphi=-3\pi^{(s)}/k^{2},
\end{eqnarray}
where we defined $u\equiv v/k$.
Combining Eqs. (43) \& (44) of [4] we obtain 
that $\delta_{P_{M}}=w_{M}\delta_{\rho_{M}}$. 
The perturbed Raychaudhuri equation (Eq. 42 of [4]) is given by
\begin{eqnarray}
&&-\varphi''+\mathcal{Q}(\alpha'-\varphi')+(2\mathcal{Q}'-k^{2}/3)\alpha\nonumber\\
&=&\frac{1}{2}(\mathcal{Q}^{2}+K)(1+3w_{M})\delta_{\rho_{M}}.
\end{eqnarray}
Finally, the continuity and Euler equations (Eq. 48 \& 49 of [4], respectively) 
are,
\begin{eqnarray}
&&\delta'_{\rho_{M}}+(1+w_{M})(3\varphi'+k^{2}u)=0\\
&&u'+(1-3w_{M})\mathcal{Q}u+\frac{w'_{M}u}{1+w_{M}}\nonumber\\
&=&\alpha+\frac{w_{M}\delta_{\rho_{M}}}{1+w_{M}}
-\frac{2}{3}\left(\frac{1-3K/k^{2}}{1+w_{M}}\right)\frac{\pi^{(s)}}{\rho_{M}}.
\end{eqnarray}
Eqs. (9)-(14) summarize our linear perturbation equations.
We note that this system of six equations (not all 
are independent) governs the evolution of $\varphi$, $\alpha$, 
$v$, and $\delta_{\rho_{M}}$. Much like with the background equations, 
Eqs. (7) \& (8), the standard linear perturbation equations 
(obtained by setting $\phi=\sqrt{\frac{3}{8\pi G}}=constant$ 
and $\delta_{\phi}=0$ in Eqs. 38, 40, 41, 43, 44, 48 \& 49 in [4]) 
can be recovered by making the 
replacement $\mathcal{Q}\rightarrow\mathcal{H}$, $\alpha\rightarrow\alpha+\delta_{\phi}$, 
$\varphi\rightarrow\varphi+\delta_{\phi}$, 
$\delta_{\rho_{M}}\rightarrow\delta_{\rho_{M}}-4\delta_{\phi}$ 
and $\delta_{P_{M}}\rightarrow\delta_{P_{M}}-4\delta_{\phi}$. 
We thus see 
not only that the background field equations obtained in the framework described by 
Eq. (1) and applied to a maximally-symmetric spacetime are equivalent to the FRW 
dynamics of standard cosmology, but also their perturbations are. This establishes the 
equivalence between the two approaches.

Vector and tensor perturbations are described by Eqs. (53)-(55) 
and (58) of [4], respectively, and the equivalence between the models is 
straightforward to show in this case, provided 
that $a\rightarrow\phi$ and $\rho_{M}\rightarrow\rho_{M}/\phi^{4}$ 
and $\pi^{(v),(t)}\rightarrow\pi^{(v),(t)}/\phi^{4}$, 
where $\pi^{(v)}$ and $\pi^{(t)}$ are the stresses associated with 
vector and tensor mode perturbations, respectively. 

The Newtonian limit of Conformal scalar-tensor Gravity 
is obtained from Eqs. (9)-(14) by setting $\mathcal{Q}$, $K$, 
and $w_{M}$ to 0. In particular, Eqs. (9), (13) \& (14) are the Poisson, 
continuity, and Euler equations, respectively. 

\section{Summary}

The background field equations and their linear perturbations in maximally-symmetric 
spacetimes are described within the framework of conformal 
scalar-tensor gravity, and are found to be equivalent in form to the Friedmann equation and 
linear perturbation equations as obtained in the standard cosmological model. The 
only difference between the two formulations is that in the former $\mathcal{Q}=\phi'/\phi$ 
replaces $\mathcal{H}=a'/a$ of the latter, and shifted perturbation quantities 
involving fractional perturbations of the scalar field replace the standard 
perturbation quantities. In particular, this implies that space expansion is 
replaced by the scalar field evolution (i.e. mass evolution of fundamental 
particles) on a maximally-symmetric static background spacetime. We emphasize that 
$G$, the hallmark of gravity, is absent from the background field equations Eqs. (7)-(8). 
This is consistent with the fact that the background spacetime is characterized by a 
vanishing Weyl tensor and thus the background evolution on cosmological scales is 
an inertial rather than gravitational phenomenon [1].

Our reformulation of the background field equations 
and linear perturbations reproduces the standard results for the 
observables associated with 
cosmic evolution, e.g. cosmological redshift, cosmological distances, 
linear metric perturbations; likewise, 
the evolution of gravitationally bound 
objects (dubbed `halos' in the cosmological context), 
is {\it identical} to that of standard cosmology, but on 
a static rather than evolving background space (and with evolving particle masses).
Allowing for particle masses to evolve generally results in energy-momentum 
non-conservation, yet we reproduce the continuity equation and its 
perturbation (Eq. 14) for the {\it rescaled} energy density $\rho_{M}/\phi^{4}$.
In particular, the observed cosmological redshift is explained on a static 
background spacetime (with non-evolving wavelength) 
by evolving masses, i.e. Rydberg `constant'.
The initial {\it curvature} singularity afflicting the standard expanding FRW spacetime 
at $\eta=0$ is replaced by vanishing particle masses at 
that time. 

\section*{Acknowledgments}
The author is indebted to Yoel Rephaeli for numerous constructive, critical, and 
thought-provoking discussions which were invaluable for this work.
This work has been supported by a grant from the the JCF (San Diego, CA).

\end{document}